\documentclass[aps,twocolumn,prb]{revtex4-1}
\usepackage{amsfonts}
\usepackage[final,hiresbb]{graphicx}
\usepackage{amsmath}
\usepackage{amssymb}
\usepackage{amstext}
\usepackage{braket}
\usepackage{graphicx,epsf,epsfig}
\usepackage{dcolumn}
\usepackage{bm}
\usepackage{color}
\usepackage{ulem}
\usepackage{subfigure}
\usepackage{upgreek}

\usepackage{braket}
\usepackage[mathscr]{euscript}
\usepackage{amsmath}
\usepackage{amssymb}
\usepackage{amstext}

\definecolor{Green}{RGB}{0,204,102}
\definecolor{Purple}{RGB}{102,0,255}
\definecolor{Blue}{RGB}{51,153,255}
\definecolor{Red}{RGB}{151,010,010}

\begin{document}

\title{Angular momentum transport with twisted exciton wave packets}
\author{Xiaoning Zang and Mark T. Lusk}
\email{mlusk@mines.edu}
\affiliation{Department of Physics, Colorado School of Mines, Golden, Colorado 80401, USA}

\begin{abstract}
A chain of cofacial molecules with $C_N$ or $C_{Nh}$ symmetry supports excitonic states with a screw-like structure. These can be quantified with the combination of an axial wavenumber and an azimuthal winding number. Combinations of these states can be used to construct excitonic wave packets that spiral down the chain with well-determined linear and angular momenta. These twisted exciton wave packets can be created and annihilated using laser pulses, and their angular momentum can be optically modified during transit. This allows for the creation of opto-excitonic circuits in which information, encoded in the angular momentum of light, is converted into excitonic wave packets that can be manipulated, transported, and then re-emitted.   A tight-binding paradigm is used to demonstrate the key ideas. The approach is then extended to quantify the evolution of twisted exciton wave packets in a many-body, multi-level time-domain density functional theory setting. In both settings, numerical methods are developed that allow the site-to-site transfer of angular momentum to be quantified. 
\end{abstract}

 
\maketitle


The conversion of light into excitons allows information and energy to be readily manipulated, transported, and re-emitted as photons\cite{Zang_PRB_2017, Zang_2017a}. This provides a conceptual framework for creating opto-excitonic circuits that process energy and information for storage~\cite{LundstromScience1999, WinbowNL2007}, manipulation using electrostatic fields\cite{HagnAPL1995, GradientAPL2006}, gating architectures\cite{Lusk_Fano_2015}, excitonic transistors\cite{High2007, HighScience2008, warburton2008}, lattice-based exciton conveyors\cite{ConveyerPRL2011}, and directed-transfer devices\cite{DirectedAPL2010, DirectedACSnano2015}. 

In a parallel but unrelated front of research, twisted light~\cite{AllenPRA1992} has received a great deal of attention in optical computing because its photonic angular momentum (PAM) can serve as an extra degree of freedom for carrying information. Potential applications include classical data transfer \cite{ClassComOE2004, ClassComNPho2012, ClassComScience2013, ClassComPNAS2016}, quantum key distribution\cite{QComPRL2014, KeyPRA2013, KeyNJP2015}, quantum entanglement\cite{EntangleNature2001, CryptographyPRL2002, EntanglePRL2005, EntangleScience2010, EntanglePRA2012, EntangleNC2014, VectorVotexPRA2016}, and quantum cloning\cite{CloningNPho2009}.  Beams of twisted light with over 10,000 angular momentum quanta have now been experimentally realized\cite{LargePTCPNAS2016}.  They can be generated by a variety of means, such as spatial light modulators\cite{AndreyOL2013, HeckenbergOL1992}, spiral phase plates\cite{BeijersbergenOC1994}, q-plates\cite{MarrucciPRL2006, STOCPRL2009, STOCJOpt2011, STOClcdropletPRL2009}, homogeneous uniaxial birefringent crystals\cite{STOCJOSAA2003, STOCPRE2003, STOCuniaxialOL2009, STOCuniaxialOL2010, STOCuniaxialOE2010},  molecular assemblies~\cite{AndrewsPRL2013, AndrewsPRA2014} and metasurfaces\cite{STOCmetasurfAPL2014, STOCmetasurfSciRep2016, STOCmetasurfNanoLett2016}. 

\begin{figure}[hptb]
\begin{center}
\includegraphics[width=0.45\textwidth]{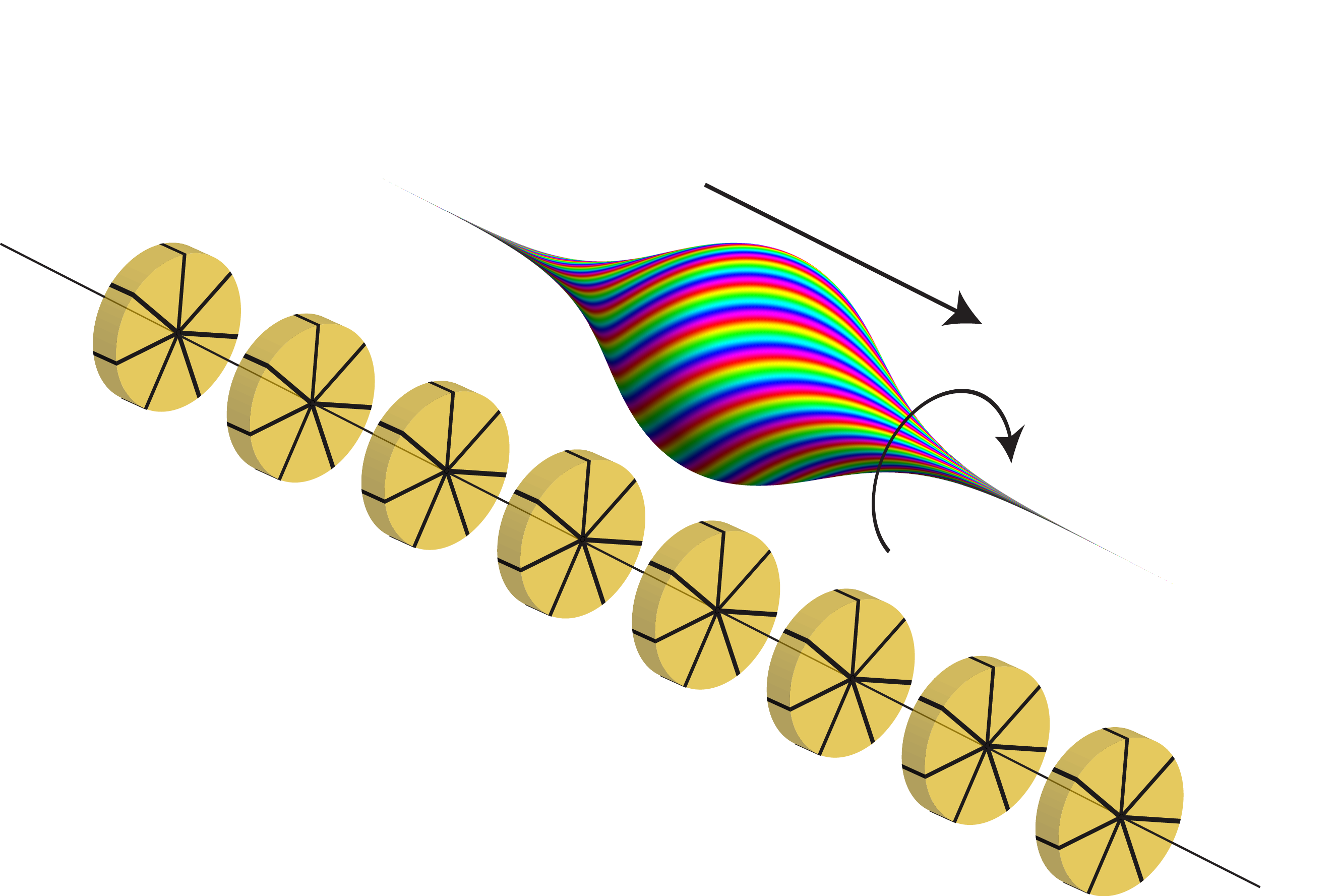}
\end{center} 	
\caption{\emph{Twisted exciton wave packets.} A TE wave packet travels down a chain of molecules with $C_7$ or $C_{7h}$ symmetry. Colors represent the evolving phase of the quantum amplitudes, here depicted as a continuum although these are discrete in practice.}\label{TXphasefront}
\end{figure}

Intriguingly, it may be possible to combine the advantages of converting light into excitons with the ability of light to encode information as angular momentum. In the case of a single molecule, we have  shown that photons with angular momentum can be converted into \emph{twisted excitons} (TE) defined in terms of an excitonic angular momentum (EAM)~\cite{Zang_2017a}.  Multiple absorption events can be used to create a wide range of EAM, and subsequent emission produces light that exhibit this same range. This raises the question of whether or not TE wave packets, illustrated in Fig. \ref{TXphasefront}, can be laser-generated on a chain of such molecules that then serve as conduits for the transport of information. If so, the packets could be created with prescribed shape and speed, with quantum interference then used as a handle to manipulate the information that they carry~\cite{Zang_PRB_2017}. 
%
%
%

A combination of theory and computational simulation is used to show that it is indeed possible to construct pulses of twisted light that generate the desired TE packets and that the speed, footprint and EAM of these packets can be tailored. Simulations are also used to demonstrate that the EAM of a packet can be changed using a second light pulse. A tight-binding (TB) setting is first used to show how this works, and the idea is subsequently generalized to the more realistic many-body, multi-energy-level setting offered by real-time time-domain density functional theory (RT-TD-DFT).

\section{Approach}

\subsection{Tight-binding paradigm}
The key features of twisted exciton wave packets can be captured in a simple TB formalism. A chain of $L$ identical molecules, each with $N$ arms, are evenly spaced along either a finite line or on a ring. The molecules are assumed to have $C_N$ or $C_{Nh}$ symmetry, and each arm can support a ground state and one excited state. A semi-classical setting is adopted in which external electric fields are taken to be classical and internal Coulombic interactions are expressed in terms of hopping coefficients. The enforcement of periodic boundary conditions make the setting particularly simple and facilitates an examination of the spatially extended eigenstates. This is later exchanged for a chain of finite length to  investigate laser-induced wave packets. In the absence of external light-matter interactions, the Hamiltonian is taken to be
\begin{eqnarray}
\hat{H_0}  &=& \hat{H}_\Delta  + \hat{H}_{\rm arm} +  \hat{H}_{\rm chain}, \nonumber \\
\hat{H}_\Delta  &=& \sum_{n=1}^{L}\sum_{j=1}^{N}\Delta \hat{c}_{nj}^{\dagger}\hat{c}_{nj}, \nonumber \\
\hat{H}_{\rm arm}   &=& \sum_{n=1}^{L}\sum_{\braket{i,j}}^{N}\tau_{\rm arm}\hat{c}_{nj}^{\dagger}\hat{c}_{ni} + H.c. ,  \\  
\hat{H}_{\rm chain}   &=& \sum_{\braket{m,n}}^{L} \sum_{j=1}^{N}\tau_{\rm chain}\hat{c}_{mj}^{\dagger}\hat{c}_{nj} + H.c.    \nonumber \\
\end{eqnarray}

Here $\Delta$ is the excited state energy of each arm, $\tau_{\rm arm}$ is the coupling between nearest arms of a given molecule, $\tau_{\rm chain}$ is the coupling between the neighboring arms on adjacent molecules,  $\hat{c}_{nj}^{\dagger}$ is the creation operator for arm $j$ on molecule $n$, and $\braket{i,j}$ denotes nearest neighbors $i$ and $j$.

In an extension of work associated with single molecules~\cite{AndrewsPRL2013}, it is straightforward to show that the ground state is $\ket{0}=\prod_{n=1}^{L}\prod_{j=1}^{N}\ket{\xi^{nj,0}}$ while the $N\times L$ excited states are
\begin{equation}
\ket{v_{q_e}^{(k)}}=\sum_{n=1}^{L}\sum_{j=1}^{N}\frac{\varepsilon_N^{(j-1)q_e} \varepsilon_L^{(n-1)k_e}}{\sqrt{N L}}\ket{e_{nj}},
\label{TBstates_trans}
\end{equation}
with $\varepsilon_N=\mathrm{e}^{\imath 2\pi/N}$, $\varepsilon_L=\mathrm{e}^{\imath 2\pi/L}$, and  $\ket{e_{nj}}=\ket{\xi^{nj,1}}\prod_{m\neq n}^{L}\prod_{i\neq j}^{N}\ket{\xi^{mi,0}}$. The EAM, $q_e$, is an integer bounded by $\frac{-1}{2}(N+1)$ and $\frac{1}{2}(N-1)$. The non-dimensional wave number, $k_e$, is the analogous characterization of linear momentum down the chain, an integer ranging from $\frac{-1}{2}(L+1)$ to $\frac{1}{2}(L-1)$. The  corresponding energies are 

\begin{equation}
\mathbb{E}_{q_e, k_e} = \Delta + 2\tau_{\rm arm} \cos\biggl(\frac{2\pi q_e}{N}\biggr) +2\tau_{\rm chain} \cos\biggl(\frac{2\pi k_e}{L}\biggr)
\label{TBenergies_trans}
\end{equation}
where a hollow $\mathbb{E}$ is used to distinguish exciton energy from electric field, $E$.

Now introduce semi-classical light-matter coupling via two Hamiltonians: $\hat H_1$ which governs light-mediated interactions between the ground state and each molecular eigenstate; and $\hat H_2$ which governs the analogous laser interactions that can cause transitions between eigenstates. The angular momentum of incident electric fields may be manifested as a circular polarization, a vector vortex, or linear polarization with a scalar vortex, but we restrict attention to the first two types. An electric dipole approximation is made for each arm, and the discrete rotational symmetry ensures that a rotation of the molecule about its axis by $2\pi/N$ maps one dipole into the next. It is assumed that the wavelength of the laser is much larger than the length of the system so that its spatial dependence can be dropped. Under these conditions, the details of electric field structure and dipole orientations are irrelevant, and the light-matter interactions are well-captured by the following Hamiltonians, which are functions of the PAM of the incident light, $q_p$, and the phase shift between neighboring molecules, $k_p$:
\begin{eqnarray}
&&\hat H_1(q_p, k_p) = -\mu_{0}^* E  \sum_{n=1}^{L}\sum_{j=1}^{N} \varepsilon_N^{-q_p(j-1)} \varepsilon_L^{-k_p(n-1)} \hat{c}_{nj}^{\dagger}\hat{c}_0 + H.c. \nonumber \\
&&\hat H_2(q_p, k_p) =  -\mu_{{\rm arm}}^* E \sum_{n=1}^{L}\sum_{j=1}^{N} \varepsilon_N^{-q_p(j-1)} \hat{c}_{n, {\rm mod}(j, N)+1}^{\dagger}\hat{c}_{nj}  \nonumber\\
&& - \mu_{{\rm chain}}^* E \sum_{n=1}^{L}\sum_{j=1}^{N} \varepsilon_L^{-k_p(n-1)} \hat{c}_{{\rm mod}(n, L)+1,j}^{\dagger}\hat{c}_{nj} + H.c.
\label{H12_trans}
\end{eqnarray}
The ${\rm mod}(\cdot, N)$ function returns its argument modulo $N$ and use has been made of the fact that $\varepsilon^{-q_e({\rm mod}(j, N)-1)} = \varepsilon^{-q_e(j-1)}$. The scalars, $\mu^*_0 E$,  $\mu_{{\rm arm}}^* E$, and $\mu_{{\rm chain}}^* E$ represent the inner product of electric transition dipole moments with a time-dependent electric field.

The total Hamiltonian, $\hat H(q_p, k_p) = \hat H_0 + \hat H_1(q_p, k_p) + \hat H_2(q_p, k_p)$, is then applied to the Schr{\" o}dinger equation with solutions assumed to be of the form
\begin{equation}
\ket{\Psi(t)} = A_0(t) \ket{0} + \sum_{n=1}^L \sum_{j=1}^N A_{nj}(t) \ket{e_{nj}}.
\label{LCAO_trans}
\end{equation}
This results in a set of $(N\times L)+1$ coupled ordinary differential equations that can be solved numerically for a prescribed electric field and initial state. The evolving state can then be projected onto each excitonic eigenstate to determine their population as a function of time:
\begin{eqnarray}
\rho_{q_e}^{(k_e)}(t) &=& |\braket{\Psi(t),v(q_e, k_e)}|^2 \nonumber \\
&=& \biggl(\sum_{n=1}^{N}  A^*_{nj}(t) \frac{\varepsilon_N^{(j-1)q_e} \varepsilon_L^{(n-1)k_e} }{\sqrt{N L}}\biggr)^2 .
\label{pop_trans}
\end{eqnarray}
%

\subsection{Real-time time-domain density functional theory}

The computational paradigm of RT-TD-DFT allows the time evolution of excited electronic states to be modeled explicitly~\cite{RGtddftPRL1984}. Multiple energy levels, an accounting of many-body interactions, and the ability to engineer laser pulses lends itself to a study of TE dynamics on molecular chains. With this setting, the time evolving, many-body Schr{\" o}dinger equation can be re-cast into an approximate, computationally tractable form via a standard Kohn-Sham reformulation:
\begin{eqnarray}
\imath\frac{\partial}{\partial t}\psi_i(r,t)&=&\Big[ -\frac{1}{2}\Delta^2 +\nu_{ext}(r,t)+\nu_{Ha}[\rho](r,t) \nonumber\\
&&+\nu_{xc}[\rho](r,t)\Big]\psi_i(r,t)\\
\rho(r,t)&=&2\sum_i^{N}|\psi_i(r,t)|^2 \label{dens_trans}.
\end{eqnarray}
Here $\nu_{ext}$ is the external potential which includes the potential from nuclei and an extra laser field potential. The Hartree, $\nu_{Ha}$, and exchange-correlation potential, $\nu_{xc}$, both depend on electron density. Eq. (\ref{dens_trans}) is the spin-reduced electron density and $2N$ is the number of electrons.

The time-propagated KS orbitals, $\psi_i(t)$, with $i=1, \cdots, N$, can each be expanded in the basis of the ground state KS orbitals $\{\phi_i=\psi_i(0)\}$. The time-propagated multi-electron wave function can thus be represented as\cite{TimePropState}
\begin{eqnarray}
\Psi(t)&=&\ket{\psi_1(t)\psi_2(t)\cdots\psi_N(t)}\nonumber\\
&=&\sum_{i\ne j\ne\cdots\ne l}^N C_{ij\cdots l}(t)\ket{\phi_i\phi_j\cdots\phi_l}.
\label{TPwf_trans}
\end{eqnarray}
The multi-electron coefficients are $C_{ij\cdots l}(t) = c_{1i}(t)c_{2j}(t)\cdots c_{Nl}(t)$ with $c_{mn}(t)=\braket{\phi_n|\psi_m(t)}$. The coefficients of each determinant $\Psi_a^i$ are straightforward to obtain:
\begin{equation}
c_a^i(t)=\braket{\Psi_a^i|\Psi(t)}.
\label{ct}
\end{equation}
Here $\Psi_a^i=\ket{\phi_1\cdots\phi_i\cdots\phi_N}$ means one electron is excited from the $a^{th}$ occupied KS orbital to the $i^{th}$ unoccupied KS orbital.

This modeling strategy was implemented using the computational package OCTOPUS~\cite{OCTOPUS}. Simulation domains amount to a collection of spheres created around each atom that have a radius of $5.67\, \rm{Bohr}$. A spatial grid of $0.284\, \rm{Bohr}$ was used to discretize this domain. A generalized gradient approximation (GGA) parametrized by Perdew, Burke, and Ernzerhof (PBE)\cite{PBE} was used to account for exchange and correlation effects, and a Troullier Martins pseudopotential was employed~\cite{Troullier_1991}. The simulation time step was set to $0.027\, \rm{a.u.}$ 

Although it is possible to run simulations for which ions move, they were frozen for the sake of computational expediency. Electron-photon interactions, and decoherence in particular, were therefore not considered. Even within this setting, using a well-parallelized high-performance computing platform, it is computationally challenging to carry out real-time, multi-electron dynamics. This motivated the construction of molecular sites in which each arm is a hydrogen dimer and three arms were considered per site (Fig. \ref{Dimer_Triad}(a)). It is then possible to track the total excited state population on each site as a function of time using a previously developed methodology~\cite{Zang_PRB_2017}. For two-site systems, though, it is also possible to track the evolution of each angular momentum component on each site, and an approach for doing this is taken up next.

\subsubsection{Evolution of EAM components on a single site}

We first introduce a method for calculating the EAM population on a single, three-arm site. Consider the absorption of a laser pulse with spin $\pm1$ by a three-arm $H_2$ system (Fig. \ref{Dimer_Triad}(a)). The right side of Eq. (\ref{TPwf_trans}) will be an excited state with $EAM=\pm1_e$. This is a linear combination of $\ket{\Psi_3^4}$ and $\ket{\Psi_2^4}$:
\begin{equation}
c_2^4(t)\ket{\Psi_2^4}+c_3^4(t)\ket{\Psi_3^4}=\ket{\phi_1,c_3^4(t)\phi_2-c_2^4(t)\phi_3,\phi_4}.
\label{EAMpm1}\end{equation}
The ground state is symmetric on all arms and therefore its EAM is $0_e$. The difference between the right side of Eq. (\ref{EAMpm1}) and the ground state determinant is that KS orbitals $\phi_2$ and $\phi_3$ are replaced by $c_3^4(t)\phi_2 - c_2^4(t)\phi_3$ and $\phi_4$. It is this replacement that is the origin of the observed time variation in the quantum amplitude phase of each arm and, therefore, the computational realization of a non-zero EAM. However the KS orbital $\phi_4$ is symmetric on all arms, which means it actually only $c_3^4(t)\phi_2-c_2^4(t)\phi_3$ that is responsible for the EAM. We can therefore expand $c_3^4(t)\phi_2-c_2^4(t)\phi_3$ in the basis of $\{\ket{e_j}\}$ and take the following projection to get population of $EAM=\pm1$ states:
\begin{eqnarray}
P_{+1_e}&=&2|\braket{v_{1}|c_3^4(t)\phi_2-c_2^4(t)\phi_3}|^2\nonumber\\
P_{-1_e}&=&2|\braket{v_{-1}|c_3^4(t)\phi_2-c_2^4(t)\phi_3}|^2 .
\label{population_trans}
\end{eqnarray}
Here $\ket{v_{\pm1}}$ is the excited state with $EAM=\pm1$ from Eq. (\ref{TBstates_trans}) with $L = 1$ and $N = 3$, and the factor $2$ accounts for the fact that spin can be either up or down.

\subsubsection{Evolution of EAM components on a pair of sites}

\begin{figure}[hptb]
\begin{center}
\includegraphics[width=0.45\textwidth]{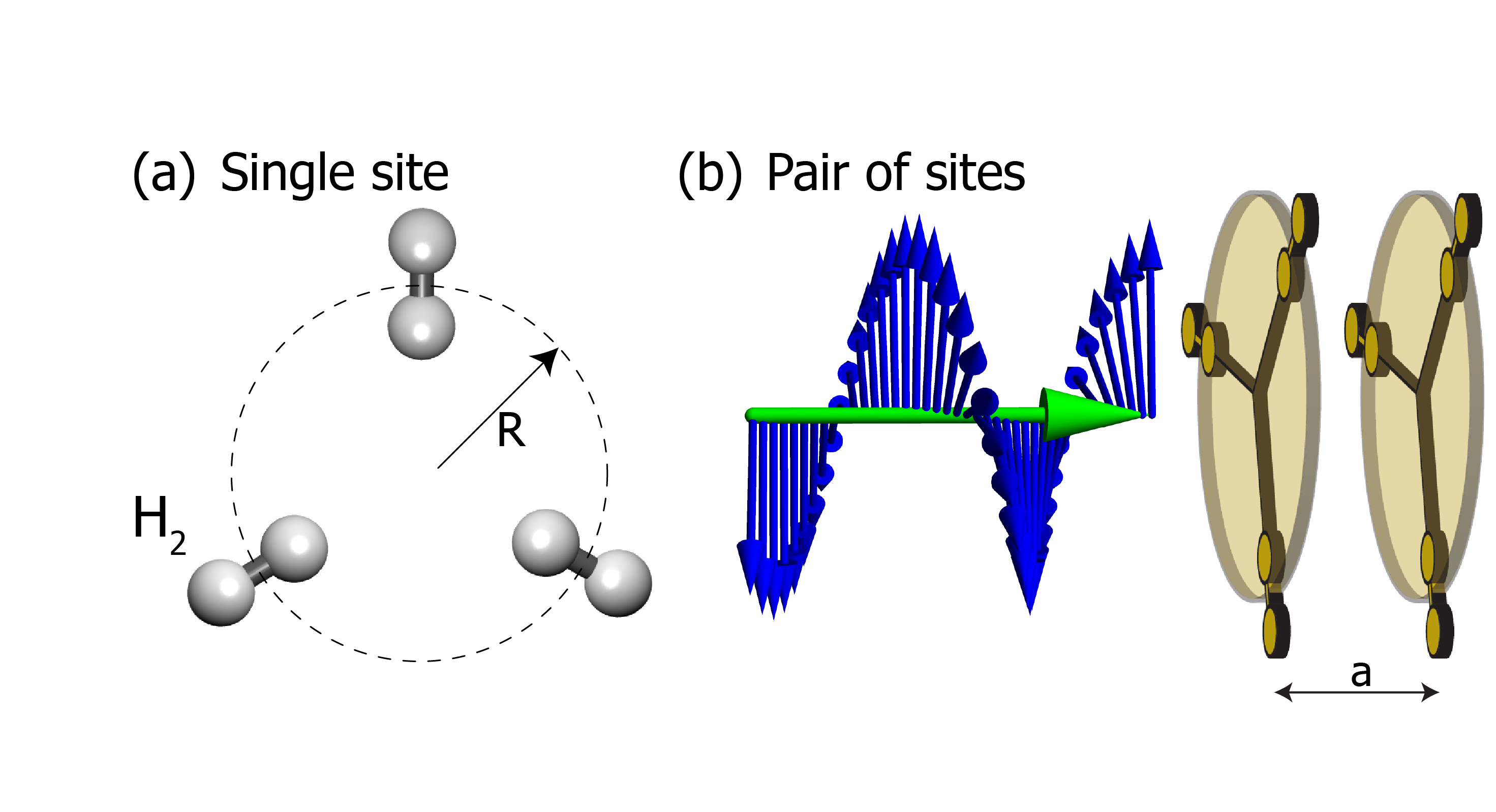}
\end{center}
\caption{\emph{Three-arm $H_2$ sites.}  (a) A single site is composed a triad of $H_2$ dimers that lie in the plane of a disk of radius, $R$. Each $H_2$ has a bond length of $1.40\,\rm{Bohr}$. (b) Laser excitation from the left induces a Rabi oscillation in a pair of sites.}
\label{Dimer_Triad}
\end{figure}

The single-site result of Eq. (\ref{population_trans}) is now extended to track the EAM components on each molecule of a two-site system. Consider the absorption of a laser pulse with spin $1$ by the left member of two-site, three-arm $H_2$ molecular system with $R=3.78\, \mathrm{Bohr}$ (Fig. \ref{Dimer_Triad}(b)). The right side of Eq.(\ref{TPwf_trans}) will be an excited state with $EAM=1_e$. With only dominant determinants considered, this is shown to be
\begin{eqnarray}
\Psi(t)&=&c_3^7(t)\Psi_3^7 + c_3^8(t)\Psi_3^8 + c_4^7(t)\Psi_4^7 + c_4^8(t)\Psi_4^8\nonumber\\
&+& c_5^7(t)\Psi_5^7 + c_5^8(t)\Psi_5^8 + c_6^7(t)\Psi_6^7 + c_6^8(t)\Psi_6^8.
\label{TPwf2sites}\end{eqnarray}
Eq. (\ref{TPwf2sites}) can now be simplified to show a clear exciton transport between the two sites. The KS orbitals can be constructed from orbitals associated with each site:
\begin{eqnarray}
\phi_3=\frac{1}{\sqrt{2}}(\phi_3^1+\phi_3^2), &&\quad \phi_4=\frac{1}{\sqrt{2}}(\phi_4^1+\phi_4^2)\nonumber\\
\phi_5=\frac{1}{\sqrt{2}}(\phi_3^1-\phi_3^2), &&\quad \phi_6=\frac{1}{\sqrt{2}}(\phi_4^1-\phi_4^2)\nonumber\\
\phi_7=\frac{1}{\sqrt{2}}(\phi_7^1+\phi_7^2), &&\quad \phi_8=\frac{1}{\sqrt{2}}(-\phi_7^1+\phi_7^2).
\label{decomposeKS}
\end{eqnarray}
Superscripts $1$ and $2$ denote orbitals on first and second site, respectively. RT-TD-DFT can thus be used to study the generation and evolution of individual EAM modes in response to laser excitation. 

%
%
%

Substituting Eq. (\ref{decomposeKS}) into Eq. (\ref{TPwf2sites}) and applying the fact that $c_3^7=-c_5^8$, $c_4^7=-c_6^8$, $c_5^7=-c_3^8$ and $c_6^7=-c_4^8$, the time propagated multi-electron wave function can be simplified to
\begin{equation}
\Psi(t)=\ket{\phi_7^1\psi^1(t)\phi_3^2\phi_4^2}+\ket{\phi_3^1\phi_4^1\phi_7^2\psi^2(t)}\label{decomposeWf}
\end{equation}
where
\begin{eqnarray}
\psi^1(t) &=& (c_3^7(t)-c_3^8(t))\phi_4^1-(c_4^7(t)-c_4^8(t))\phi_3^1\nonumber \\
\psi^2(t) &=& (c_3^7(t)+c_3^8(t))\phi_4^2-(c_4^7(t)+c_4^8(t))\phi_3^2.
\end{eqnarray}
The first and second terms of Eq. (\ref{decomposeWf}) are exciton states on sites $1$ and $2$, respectively.  This can be seen by simplifying the ground state so that it has the same structure:
\begin{equation}
\Psi_{gs}=\ket{\phi_1\phi_2\phi_3\phi_4\phi_5\phi_6}=\ket{\phi_3^1\phi_4^1\phi_3^2\phi_4^2}.
\label{decomposeGs}\end{equation}

In Eqs. (\ref{decomposeWf}) and (\ref{decomposeGs}), the KS orbitals, $\phi_1$ and $\phi_2$, are not shown since they are not involved in the excitation. Comparing Eq. (\ref{decomposeGs}) with the first term of Eq. (\ref{decomposeWf}), the only difference is that $\phi_3^1$ and $\phi_4^1$ are replaced by $\phi_7^1$ and $\psi^1(t)$. The orbital $\phi_7^1$ has no phase difference among three arms and therefore has no  contribution to EAM. The $\psi^1(t)$ must therefore be responsible for any nonzero EAM on the left site. By the same reasoning, $\psi^2(t)$ must be responsible for any nonzero EAM on the right site. Eq. (\ref{population_trans}) can then be used to quantify populations of EAM states on each site:
\begin{equation}
P^i_{\pm1_e}=2|\braket{v_{\pm1}|\psi^i(t)}|^2.
\label{population_2-site}
\end{equation}
%

\subsection{Laser-generation of twisted exciton wave packets}

Independent of the computational paradigm chosen, laser pulses can be designed so as to generate excitonic wave packets of prescribed linear and angular momentum.  Towards this end, consider a periodic ring system of $L$ identical molecules, each with $N$ identical arms. Each arm is assumed to support a ground state and one excited state so that any dynamical process can be represented as a linear combination of such states with time-varying coefficients. The Schr{\" o}dinger equation can be then expressed as a set of $N\times L$ coupled ordinary differential equations for the quantum amplitudes of each arm on each molecule, $u_{n,j}$:
\begin{eqnarray}
i\hbar \dot{u}_{n,j}&=&\Delta u_{n,j}+\tau_{\rm arm}u_{n,j+1} + \tau_{\rm arm}u_{n,j-1}\nonumber\\ 
&+& \tau_{\rm chain}u_{n+1,j}+\tau_{\rm chain}u_{n-1,j}.
\label{CoupRingEqs}
\end{eqnarray}
Here the subscripts $n$ and $j$ identify the molecule and the arm, respectively.

An analogous set of equations for a finite (non-periodic) chain of $L$ identical molecules with $N$ arms for each can also be constructed. As justified previously\cite{Zang_PRB_2017}, assume that only the first molecule can be excited via interaction of its transition dipole, $\vec{\mu}_0$, with an external circularly polarized or vector vortex field. The arm amplitudes, $q_{n,j}(t)$, then evolve according to the following equations:
\begin{eqnarray}
\imath\hbar\dot{q}_{0,j}&=&-\mu_0E\varepsilon^{q_e(j-1)}q_{1,j}\nonumber\\
\imath\hbar\dot{q}_{1,j}&=&-\mu_0(E\varepsilon^{q_e(j-1)})^*q_{0,j}+\Delta q_{1,j}+\tau_{\rm arm}q_{1,j-1}\nonumber\\
&&+\tau_{\rm arm}q_{1,j+1}+\tau_{\rm chain}q_{2,j}\\
\imath\hbar\dot{q}_{n,j}&=&\Delta q_{n,j}+\tau_{\rm arm}q_{n,j+1}+\tau_{\rm arm}q_{n,j-1}\nonumber\\
&&+\tau_{\rm chain}q_{n+1,j}+\tau_{\rm chain}q_{n-1,j}\nonumber\\
\imath\hbar\dot{q}_{L,j}&=&\Delta q_{L,j}+\tau_{\rm arm}q_{L,j+1}+\tau_{\rm arm}q_{L,j-1}+\tau_{\rm chain}q_{L-1,j}.\nonumber
\label{CoupChainEqs}
\end{eqnarray}
Here $1<j<N$ and $1<n<L$. The phase factor is $\varepsilon=e^{\imath 2\pi/N}$, and the ground state occupation is given by $q_{0,j} = \braket{\xi^{1,j,0}|\Psi(t)}$.

A comparison of Eqs. (\ref{CoupRingEqs}) and (\ref{CoupChainEqs}) suggests that the exciton dynamics of the ring system can be elicited on a finite chain provided functions $q_{0,j}(t)$ and $E(t)$ are chosen so as to satisfy the following conditions:
\begin{eqnarray}
i\hbar\dot{q}_{0,j}&=&-\mu_0E\varepsilon^{q_e(j-1)}u_{1,j}\nonumber\\
-\mu_0(E\varepsilon^{q_e(j-1)})^*q_{0,j}&=&\tau_{\rm chain}u_{L,j}\label{q0andE}.
\end{eqnarray}
Multiplication of the first equation by $q_{0,j}^*$ gives
\begin{equation}
\hbar\dot{q}_{0,j}q_{0,j}^*=i\mu E\varepsilon^{q_e(j-1)}q_{1,j}q_{0,j}^*\label{q0andE2}.
\end{equation}
Summation of Eq. (\ref{q0andE2}) and its complex generates
\begin{equation}
\hbar\dot{\rho}_{0,j}=-2\tau_{\rm chain}\mathrm{Im}(u_{L,j}u_{1,j}^*),\label{rhot}
\end{equation}
with $\rho_{0,j}(t)=q_{0,j}q_{0,j}^*$ and $\rho_{0,j}(t=0)=1$ as the initial condition. Then Eq. (\ref{rhot}) can be integrated to give
\begin{equation}
\rho_{0,j}(t)=1-\frac{2\tau_{\rm chain}}{\hbar}\int_0^t\mathrm{d}t_1\mathrm{Im}(u_{L,j}(t_1)u_{1,j}^*(t_1))\label{intrhot}.
\end{equation}
The ground state amplitude can be written as
\begin{equation}
q_{0,j}(t)=A_j(t)e^{i\varphi_j(t)}\label{complexq0}.
\end{equation}
Substitute Eq. (\ref{complexq0}) into first expression of Eq. (\ref{q0andE}) and multiply by $\dot{A}_j(t)$ to obtain
\begin{equation}
i\hbar(A_j\dot{A_j}+iA_j^2\dot{\varphi}_j)=\tau_{\rm chain}u^*_{L,j}u_{1,j}.
\end{equation}
This can be simplified by noting that $A_j\dot{A}_j=\frac{1}{2}\dot{\rho}_{0,j}$ and using Eq. (\ref{rhot}) so that
\begin{equation}
\dot{\varphi}_j(t)=-\frac{\tau_{\rm chain}}{\hbar\rho_{0,j}}\mathrm{Re}(u_{L,j}^*u_{1,j})\label{phit}.
\end{equation}
With the intitial condition of $\varphi_j(t=0)=0$, we therefore have
\begin{equation}
\varphi_j(t)=-\frac{\tau_{\rm chain}}{\hbar}\int_0^{t}\mathrm{d}t_1\rho_{0,j}(t_1)\mathrm{Re}(u_{L,j}^*(t_1)u_{1,j}(t_1))\label{intphit}.
\end{equation}
The ground state amplitude, Eq. (\ref{complexq0}), is completely determined. This is only an intermediate step towards the real goal of constructing an appropriate laser pulse though. We therefore substitute this expression  into second expression of Eq. (\ref{q0andE}) and take $j=1$ to get the laser pulse
\begin{equation}
E(t)=-\frac{\tau_{\rm chain}u^*_{L,1}(t)}{\mu_0\sqrt{\rho_{0,1}(t)}e^{-i\varphi_j(t)}}\label{totalE}.
\end{equation}

Note that the form of the above electric field is unphysical since it is complex valued. As previously discovered though\cite{Zang_PRB_2017},  quantum interference evanescence implies that the the following real-valued substitute gives an excellent approximation to the desired wave packet:
\begin{equation}
E_{phys}(t)=-2\mathrm{Re}\left[\left(\frac{\tau_{\rm chain}u_{L,1}(t)}{\mu_0q_{0,1}(t)}\right)^*\right] .
\label{laser_pulse}
\end{equation}
%

\section{Results}

\subsection{Tight-binding}

\begin{figure}[hptb]
\begin{center}
\includegraphics[width=0.45\textwidth]{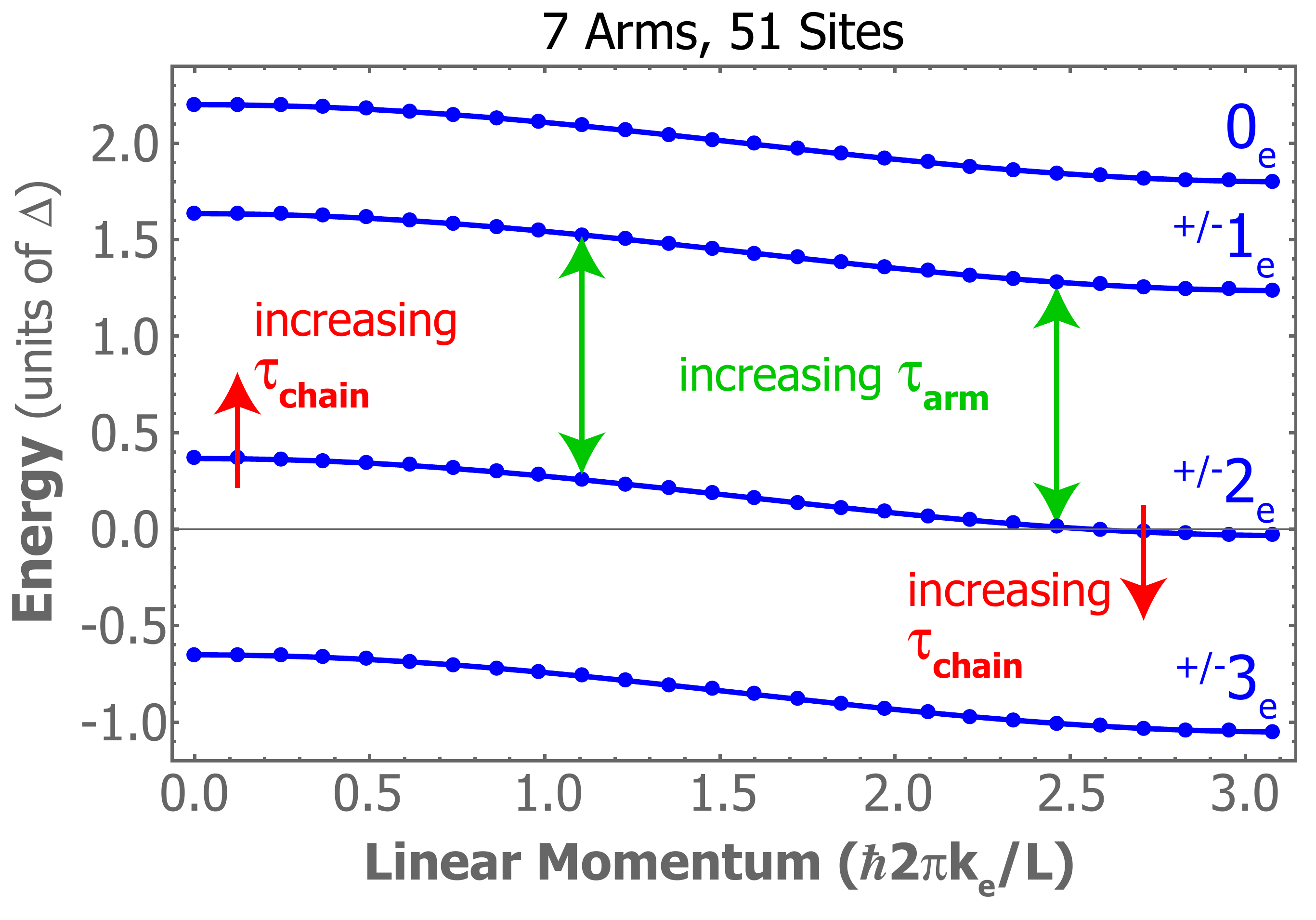}
\end{center}
\caption{\emph{Exciton band structure (tight binding).} Exciton energies as a function of linear momentum index, $k_e$, and angular momentum index, $q_e$ EAM for a 51-site chain of 7-arm molecules. Here $\tau_{\rm arm} = 0.75 \Delta$ and $\tau_{\rm chain}=0.1 \Delta$. The solid blue curves are a guide to the eye.}
\label{eigensystem}
\end{figure}

The tight-binding setting was considered first to elucidate exciton dynamics on a chain of 51 molecules that each contain seven arms. The specific parameter choices and resulting exciton band structure are given in Fig. \ref{eigensystem}. Because of the coupling between arms on neighboring molecules, $\tau_{\rm chain}$, otherwise degenerate energies spread into bands for each EAM, as shown in the figure. Their width increases with $\tau_{\rm chain}$. The coupling between arms on a given site, $\tau_{\rm arm}$, controls the energy spacing between bands. Exciton packets can be described with a linear combination of the associated eigenstates.
 
%
%
%

\begin{figure}[hptb]
\begin{center}
\includegraphics[width=0.45\textwidth]{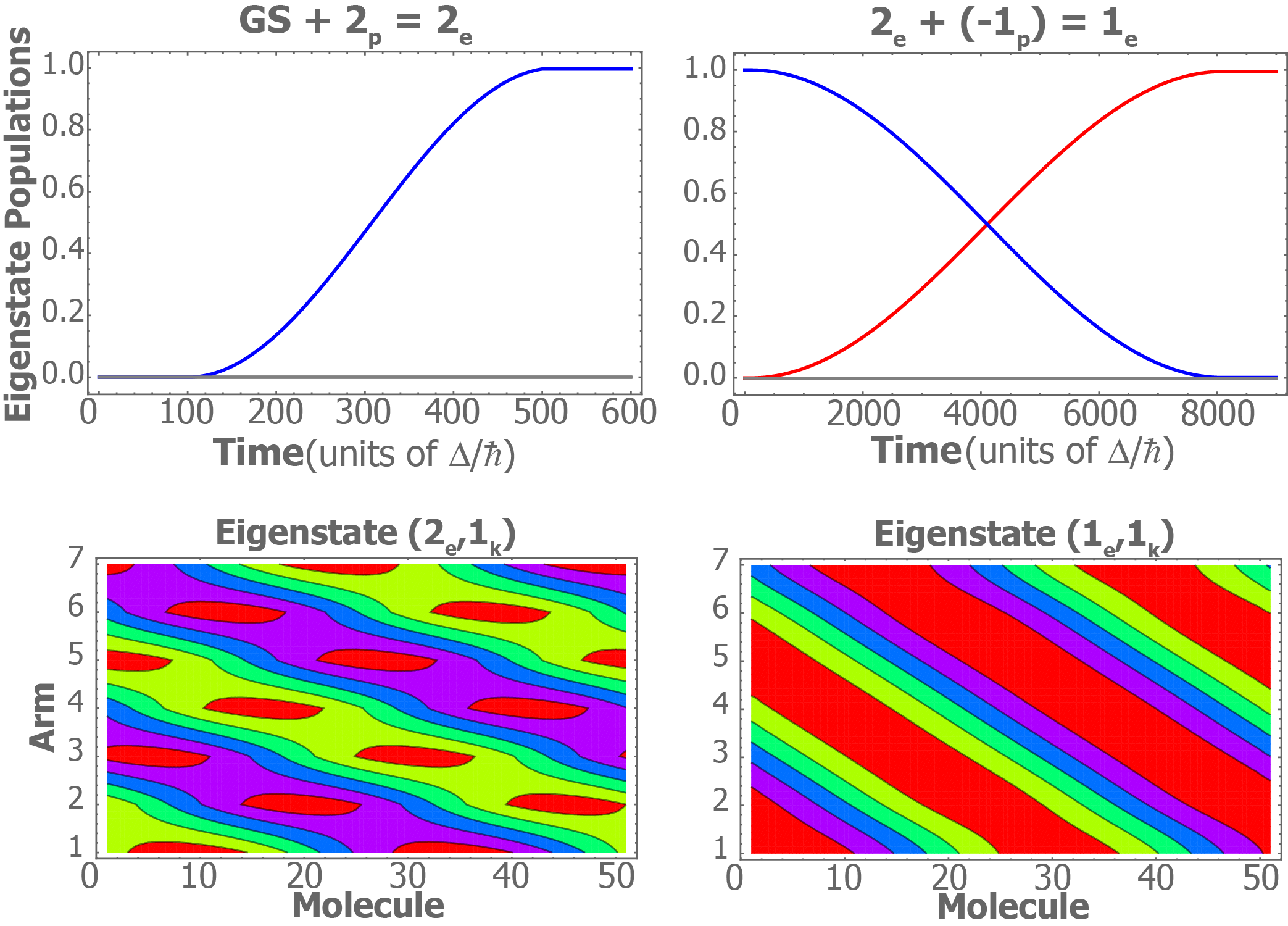}
\end{center}
\caption{\emph{Change of excitonic angular momentum on a chain (TB).} Left panels shows the AM transfer reaction (top): $GS + 2_p = 2_e$ and final excitonic state (bottom). Right panels show AM addition (top) associated with $2_e + (-1_p) = 1_e$ and final excitonic state (bottom). Colors represent the phase of the quantum amplitudes which have been linearly interpolated.}
\label{TB_Charge_Algebra}
\end{figure}

Before considering the generation of specific TE wave packets, two implementations are provided to demonstrate that a laser can be used to change the EAM on a molecular chain. If the structure is initially in its ground state, illumination of all molecules with a photonic vortex can be used to create a nonlocalized excitonic state as shown in the left panels of Fig. \ref{TB_Charge_Algebra}. There a PAM of $2_p$ is transferred to an EAM of $2_e$ using a continuous wave (CW) laser of frequency, $\omega = \mathbb{E}_{2_e, k_1}$ that is applied between $t = 100 \Delta/\hbar$ and $t = 500 \Delta/\hbar$.

Angular momentum can be subsequently withdrawn from the structure, as shown in the right panels of the same figure.  A CW laser of frequency, $\omega = \mathbb{E}_{2_e, k_1} -\mathbb{E}_{1_e, k_1}$ is applied between $t = 1000 \Delta/\hbar$ and $t = 8000 \Delta/\hbar$. Angular momentum is conserved in all such light-matter interactions within the validity of the paraxial approximation~\cite{Zang_2017a}.

%
%
%

\begin{figure}[hptb]
\begin{center}
\includegraphics[width=0.35\textwidth]{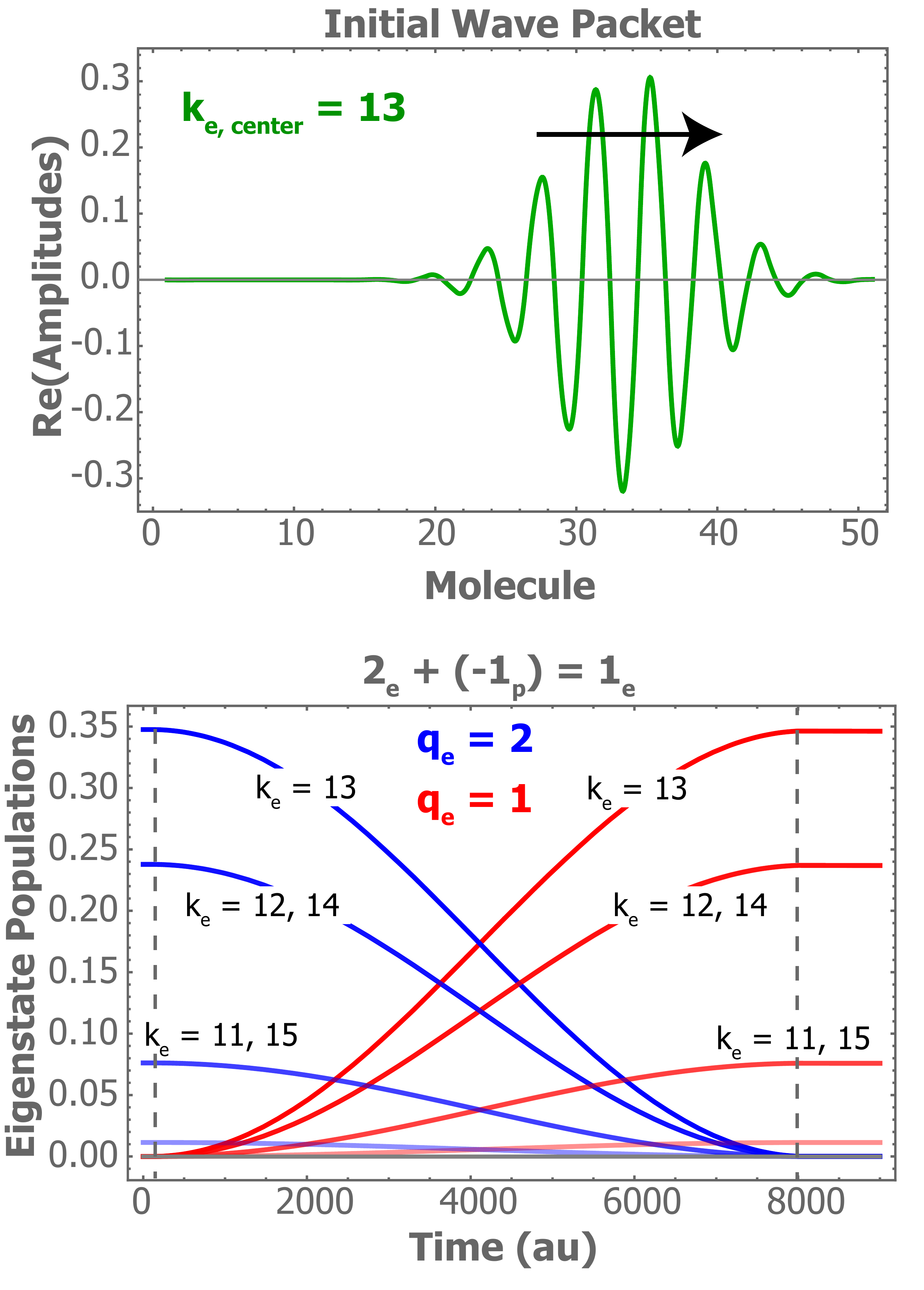}
\end{center}
\caption{\emph{Changing the EAM of a chain.} Upper panel shows initial wave packet 51-site chain of 7-arm molecules with central wave number index of $k_e = 13$. A CW laser ($-1_p$) is used to change the EAM of the packet as it moves along the ring of molecules. The curves are each associated with a particular longitudinal wave number index, $k_e$, and collectively comprise the components of the packet with EAM $= 2$ (red) and those associated with a $EAM = 1$ (blue). Parameters: $\tau_{\rm arm} = 0.75 \Delta$ and $\tau_{\rm chain}=0.1 \Delta$. The associated linear momenta are $\hbar 2\pi k_e/(L a)$ where $a$ is the longitudinal spacing between molecules and $L$ is the number of molecules.}
\label{TB_Charge_Algebra_Packet_Trans}
\end{figure}

The next step towards considering wave packets on a finite chain is to model their evolution on a periodic ring. This provides the input data required to design laser pulses to generate such packets, and it is also an ideal setting to show how EAM can be changed even while the packet travels.The initial TE wave packet with a prescribed EAM, $q_e$, and central longitudinal wavenumber, $k_0$, is used:
\begin{equation}
\ket{\Psi(0)}  = \frac{1}{\pi^{\frac{1}{4}} {(N \sigma)}^{\frac{1}{2}}}  \sum_{n,j} \mathrm{e}^{-\imath k_0 n}\mathrm{e}^{\frac{-(n-n_0)^2}{2 \sigma^2}} \varepsilon_N^{(j-1)q_e} \hat{c}^{\dagger}_{nj}\ket{\rm vac} . 
\label{initF_Transfer2}
\end{equation}
Here $\sigma$ controls the axial length of the packet, $n_0$ denotes the position of the packet center, and the length unit is the lattice spacing, $a$. As shown, in the top panel of Fig. \ref{TB_Charge_Algebra_Packet_Trans}, this packet moves to the right with a group velocity of $v(k_0) = -2 \tau_{\rm chain} \mathrm{sin}(k_0)$.

The entire system is subsequently illuminated with a windowed CW laser with a PAM $ = -1_p$. The vertical dashed lines in Fig. \ref{TB_Charge_Algebra_Packet_Trans} denote on and off times. As a result, the longitudinal wave numbers comprising the packet are unchanged, but the EAM associated with individual wave numbers transition from $EAM = 2_e$ to $EAM = 1_e$. Angular momentum is conserved, as usual, but now for a traveling wave packet. Of course, linear momentum is conserved in these light-matter interactions as well. 

%
%
%

\begin{figure}[hptb]
\begin{center}
\includegraphics[width=0.45\textwidth]{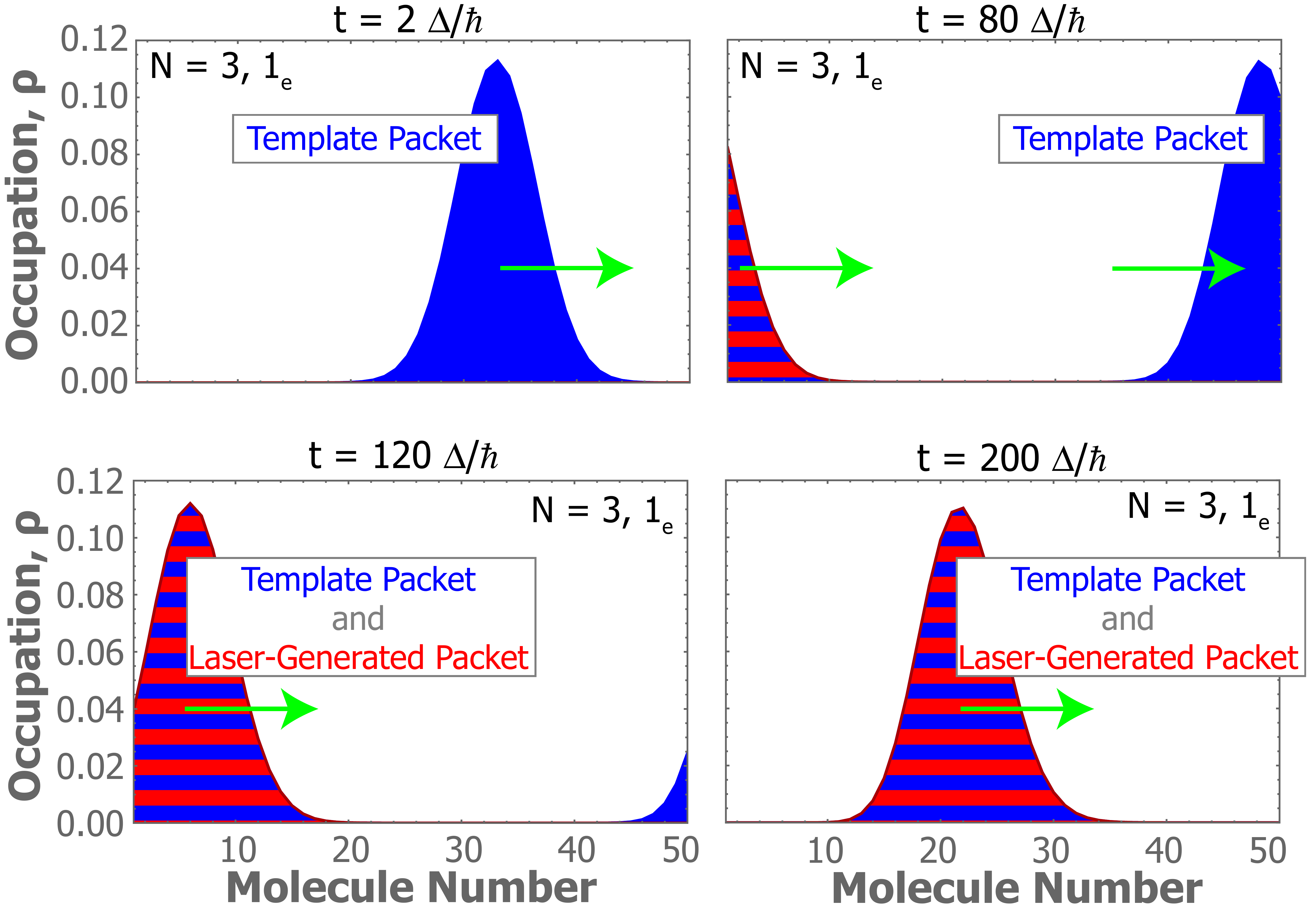}
\end{center}
\caption{\emph{Evolution of twisted exciton wave packet.} Time slices for total exciton population of each molecule on 50-site chain. Blue filled curve shows original packet on periodic ring of sites, while a red filled red curve is generated by a twisted laser pulse at left end of a finite chain. The overlap is essentially perfect, and this is depicted with a red/blue pattern. $\tau_{\rm arm} = 0.75 \Delta$.}
\label{TB_packet_evolution}
\end{figure}

\begin{figure}[hptb]
\begin{center}
\includegraphics[width=0.35\textwidth]{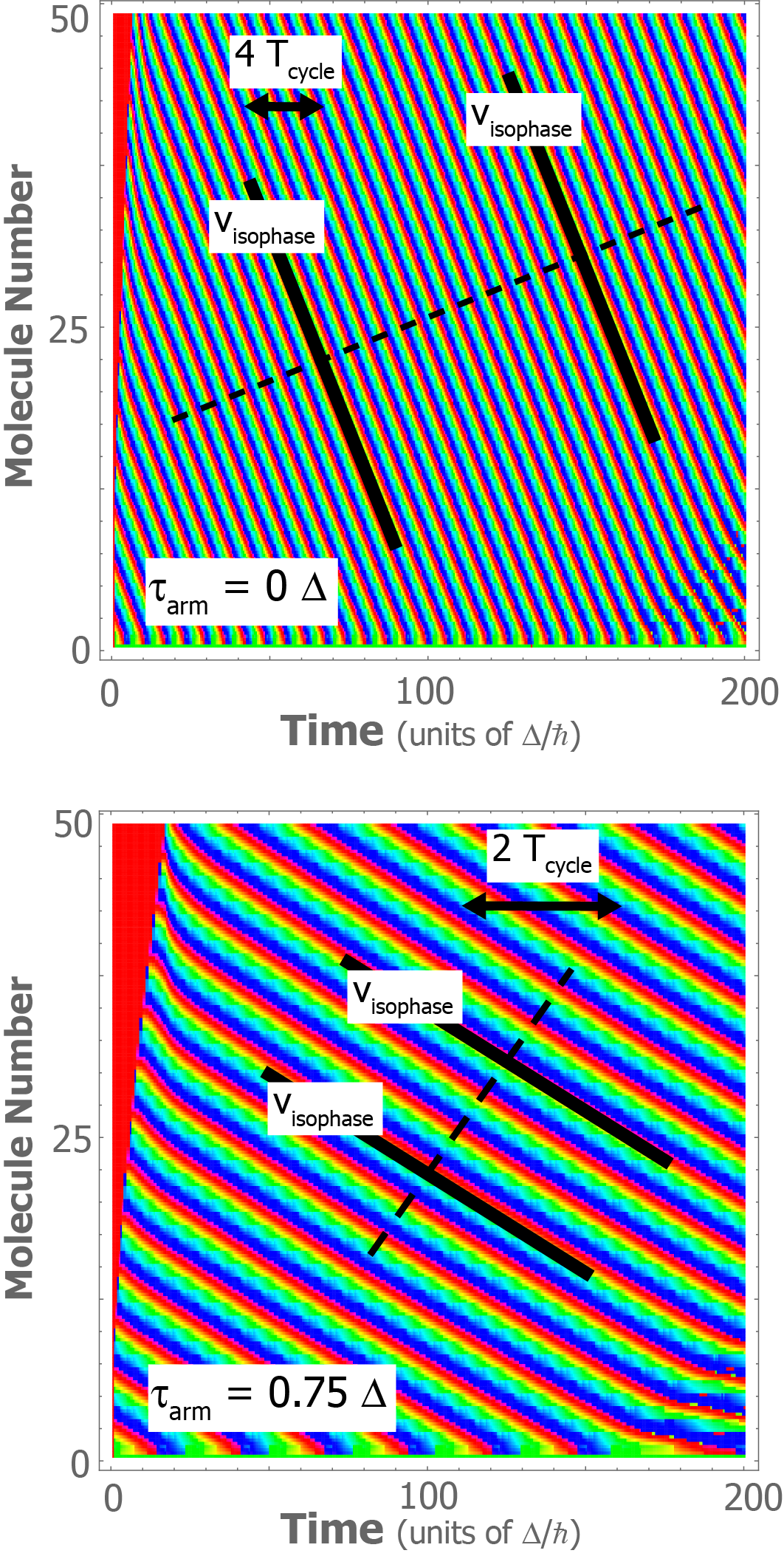}
\end{center}
\caption{\emph{Phase maps for TE wave packet.} Top panel: $\tau_{\rm arm} = 0$. Bottom panel: $\tau_{\rm arm} = 0.75 \Delta$. The thick black lines are from the theoretical predictions of Eqs. (\ref{Tcycle}) and (\ref{viso}).}
\label{Phase_Map_TB5}
\end{figure}

Wave packets with a prescribed shape, speed and EAM were next generated on a finite chain of fifty molecules, each with three arms. A laser pulse with a PAM $ = 1_p$ was applied to the chain, initially in its ground state. The evolving wave packet is shown in Fig. \ref{TB_packet_evolution}, where exciton populations on each arm of a molecule have been summed. It is clear that the twisted laser pulse has successfully generated an exciton wave packet that moves at the desired speed. The EAM of this packet can also be estimated as follows. At a given time and for a particular molecule, the phases of each arm can be used as input to determine the best fit to a rigid rotation of the associated eigenstate, $v(1_e, n_k)$. This can be carried out for all molecules and all times to produce a map showing how the phase evolves as a function of time. Two implementations, for differing value of the arm mobility, $\tau_{\rm arm}$ are shown in Fig. \ref{Phase_Map_TB5}. 

In each case, the period for a phase cycle can be calculated at a fixed site:
\begin{equation}
T_{\rm cycle} = 2 \pi/ \mathbb{E}_{q_e, k},
\label{Tcycle}
\end{equation}
where $\mathbb{E}_{q_e, k}$ is the eigenvalue given by Eq. (\ref{TBenergies_trans}). In Fig.  \ref{Phase_Map_TB5}, $q_e = 1$ and $k_{\rm center} = 13$ implying that $T_{\rm cycle} = 25.1  \Delta/\hbar$ for $\tau = 0.75 \Delta$. For comparison, $T_{cycle} = 6.28 \Delta/\hbar$ for $\tau = 0$. 

In addition, the slope of the lines of constant phase can be predicted analytically and compared with the computational results:
\begin{equation}
v_{\rm isophase} = -\frac{\lambda}{T_{\rm cycle}},
\label{viso}
\end{equation}
where $\lambda = 2 \pi/k_{\rm center}$ is the wavelength of the central frequency of the wave packet. This velocity is plotted (solid black) for both panels of Fig. \ref{Phase_Map_TB5}. This, combined with the good match in the phase cycles, $T_{\rm cycle}$, makes it clear that the laser pulse has generated wave packets that exhibit and preserve the intended EAM.
 
%
%
%

%
%
%

\subsection{Time-domain density functional theory}

The more realistic paradigm of RT-TD-DFT can also be used to study the generation and propagation of a twisted exciton wave packet. For a pair of sites, shown in Fig. \ref{Dimer_Triad}(b), Eqs. (\ref{decomposeKS}) and (\ref{population_2-site}) can then be used to explicitly track each component of EAM. This provides a detailed mechanistic understanding of the exciton dynamics. A special case of the light-matter interaction of Eqs. (\ref{H12_trans}) is adopted in which the dipole operator is written out explicitly and the laser pulse is described with a simple Gaussian amplitude instead of crafting a wave packet using Eq. (\ref{laser_pulse}):
\begin{equation}
\hat H_{laser}= (-e\vec{r}\cdot\vec{E})\, F(t)\mathrm{e}^{\mathrm{i\omega t}}.
\label{Hlaser}
\end{equation}
Here $e$ is the elementary unit of charge, $\vec{E}$ is the circularly polarized laser field, and $F(t)$ is the Gaussian envelope, $\mathrm{e}^{-(t-t_0)^2/2\tau^2}$. Circularly polarized light can be mathematically decomposed into a combination of radial and azimuthal vector vortices\cite{CircPolVortexOL2006}: 
\begin{equation}
\vec{E}=E_0(\vec{e}_x\pm\imath\vec{e}_y)=E_0\mathrm{e}^{\pm\imath \phi}(\vec{e}_r\pm\imath\vec{e}_\phi),\label{spinlight_trans}
\end{equation}
where $\{\vec{e}_x, \vec{e}_y\}$ and $\{\vec{e}_r, \vec{e}_\phi\}$ are the basis vectors in Cartesian and polar representations. The radial transition dipole orientations of each arm imply that only the radial vector vortex component $E_0\mathrm{e}^{-\imath \phi}\vec{e}_r$ can be absorbed. The circularly polarized laser with spin $\pm \hbar$ is therefore equivalent to a radial vector vortex with PAM = $\pm \hbar$.

\begin{figure}[hptb]
\begin{center}
\includegraphics[width=0.45\textwidth]{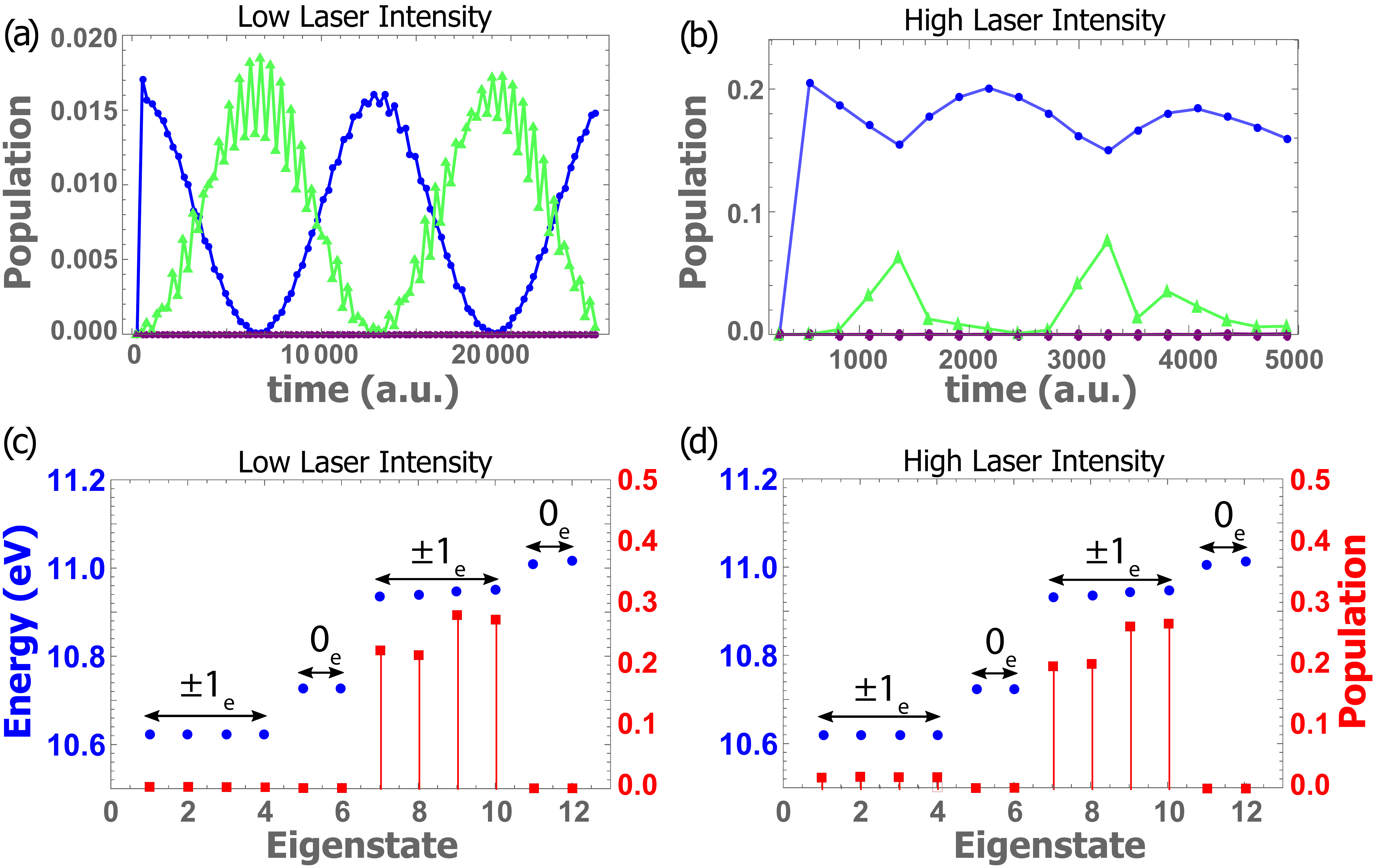}
\end{center}
\caption{\emph{Cyclic TE motion in a two-site, three-arm $H_2$ system with $R=3.78\, \mathrm{Bohr}$.} (a): Spin $+1$ laser pulses are applied to left end of the two-site system of Fig. \ref{Dimer_Triad}(b). Twisted excitons with EAM $=1_e$ are generated on the left site (blue dots) and oscillate to the right site (green triangles). Populations of EAM $=-1_e$ on both sites are denoted with purple dots. The points are connected as a guide to the eye. (c) Plot of the corresponding energy manifold and eigenstate populations. Envelop parameters are $\{t_0 = 408, \tau=81.6, \omega=0.390, E_0 = 0.002\}$ in a.u. (b, d) Same analysis but now with laser intensity that is five times greater, $E_0 = 0.010$ a.u.}
\label{DimerR2}
\end{figure}

Two sets of results are generated to show the effect of changing the arm length, $R$. Fig. \ref{DimerR2} summarizes the dynamics for $R=3.78\, \mathrm{Bohr}$, where the initially steep increase in left-site excited state shown in panels (a) and (b) is due to excitation with circularly polarized light of PAM $=1_p$. A twisted exciton with EAM $= 1_e$ is thus generated on this site. The green curve in each panel gives the population of a twisted exciton with EAM $=1_e$ on the right site, while the purple curves are the summed  populations of twisted exciton with EAM $= -1_e$ on both sites. Fig. \ref{DimerR2} (a) shows a clear Rabi oscillation albeit with a small population transfer.

In order to get a larger population, a natural thought might be to use a stronger laser. However, as shown in panel (b) of Fig. \ref{DimerR2}, such a change actually quenches the Rabi oscillation. This can be explained with the help of the eigenstate manifolds of panels (c) and (d) calculated using linear-response TD-DFT~\cite{casida1995response} with same parameters as RT-TD-DFT simulations imposed. These eigenstates can be considered as two groups. The lowest six eigenstates correspond to excited states of TB model in which only one excited state for each $H_2$ is considered. The six higher eigenstates, though, correspond to states associated with the second excited state of each $H_2$, something not accounted for in the current TB model.  The red squares are the time-averaged populations of eigenstates with the sum normalized to one. The difference in energy between the first-excited and second-excited EAM groups with $q_e= \pm 1$ is very small, and when a stronger laser pulse is applied, the eigenstates in both groups will contribute to the construction of a twisted exciton with EAM $=1_e$. This destroys the Rabi oscillation.

%
%
%

\begin{figure}[hptb]
\begin{center}
\includegraphics[width=0.45\textwidth]{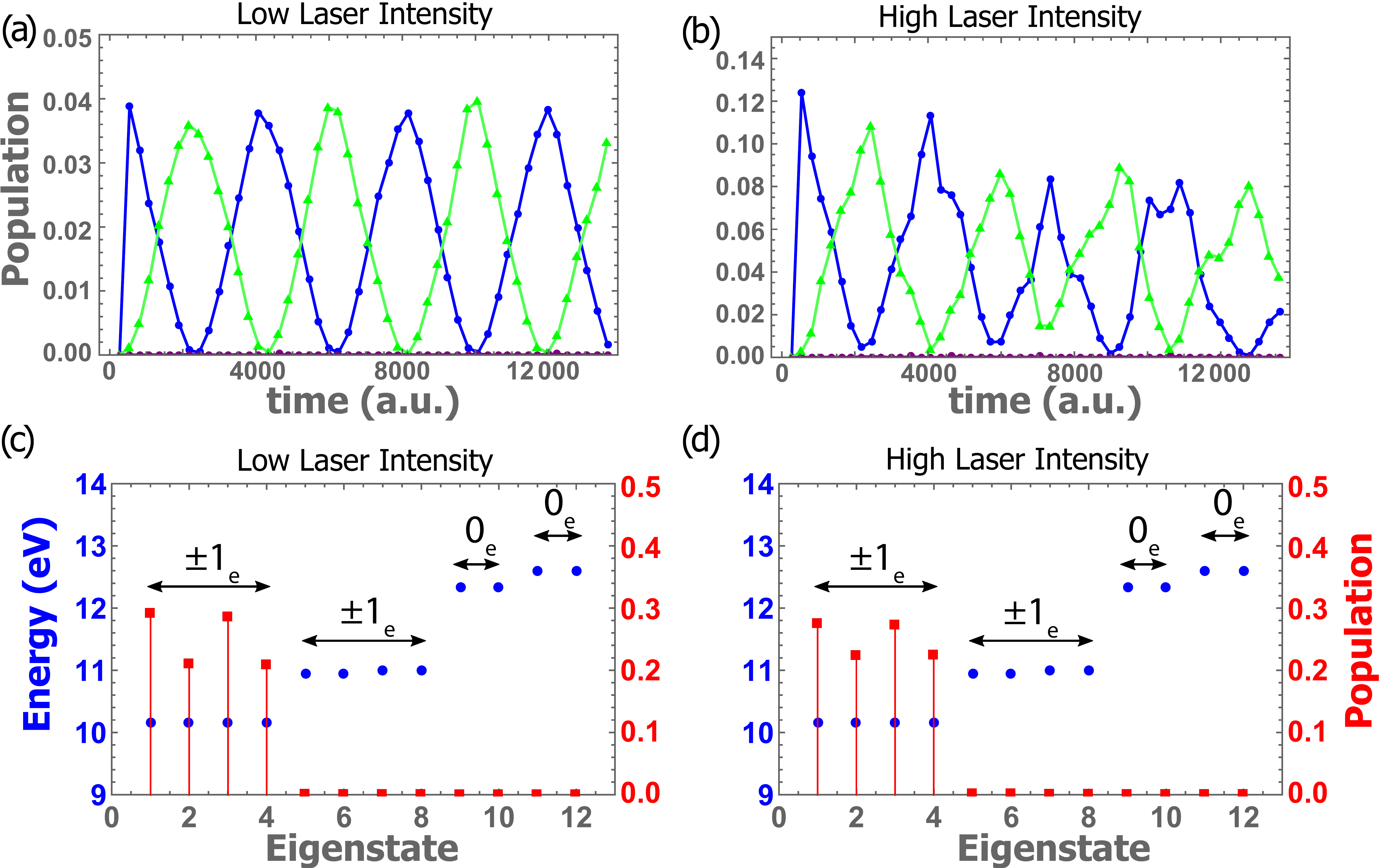}
\end{center}
\caption{\emph{TE transfers in a two-site, three-arm $H_2$ system with $R=1.89\, \mathrm{Bohr}$.} (a) and (b): a spin $1$ laser pulse is applied on one site and a twisted exciton with EAM $=1_e$ is generated on that site (blue dots). The population of EAM $=1_e$ on second site is denoted by green triangles. And the population of EAM $=-1_e$ on both sites is in purple. (c) and (d) are the corresponding energy manifold and populations of eigenstates of (a) and (b) respectively. Envelop parameters are $\{t_0 = 408, \tau=54.4, \omega=0.375, E_0 = 0.005\}$ and $\{t_0 = 408, \tau=54.4, \omega=0.375, E_0 = 0.010\}$ in a.u., respectively. }
\label{DimerR1}
\end{figure}

Systems with a large energy gap between disparate EAM groups should therefore exhibit twisted exciton Rabi oscillation with large populations. To verify this, a two-site, three-arm $H_2$ molecular system with a much smaller arm radius, $R=1.89\, \mathrm{Bohr}$, was investigated. Within the TB model, this is analogous to increasing the intra-site coupling, $\tau_{\rm arm}$, as noted in Fig. \ref{eigensystem}.  Panels (c) and (d) of Fig. \ref{DimerR1} now show a much larger energy gap between first-excited and second-excited states with the same EAM. As anticipated, a more intense laser illumination leads to larger exciton populations. As expected, only the populations of first-excited states are nonzero for both weak and strong laser illumination.

%
%
%

\begin{figure}[hptb]
\begin{center}
\includegraphics[width=0.45\textwidth]{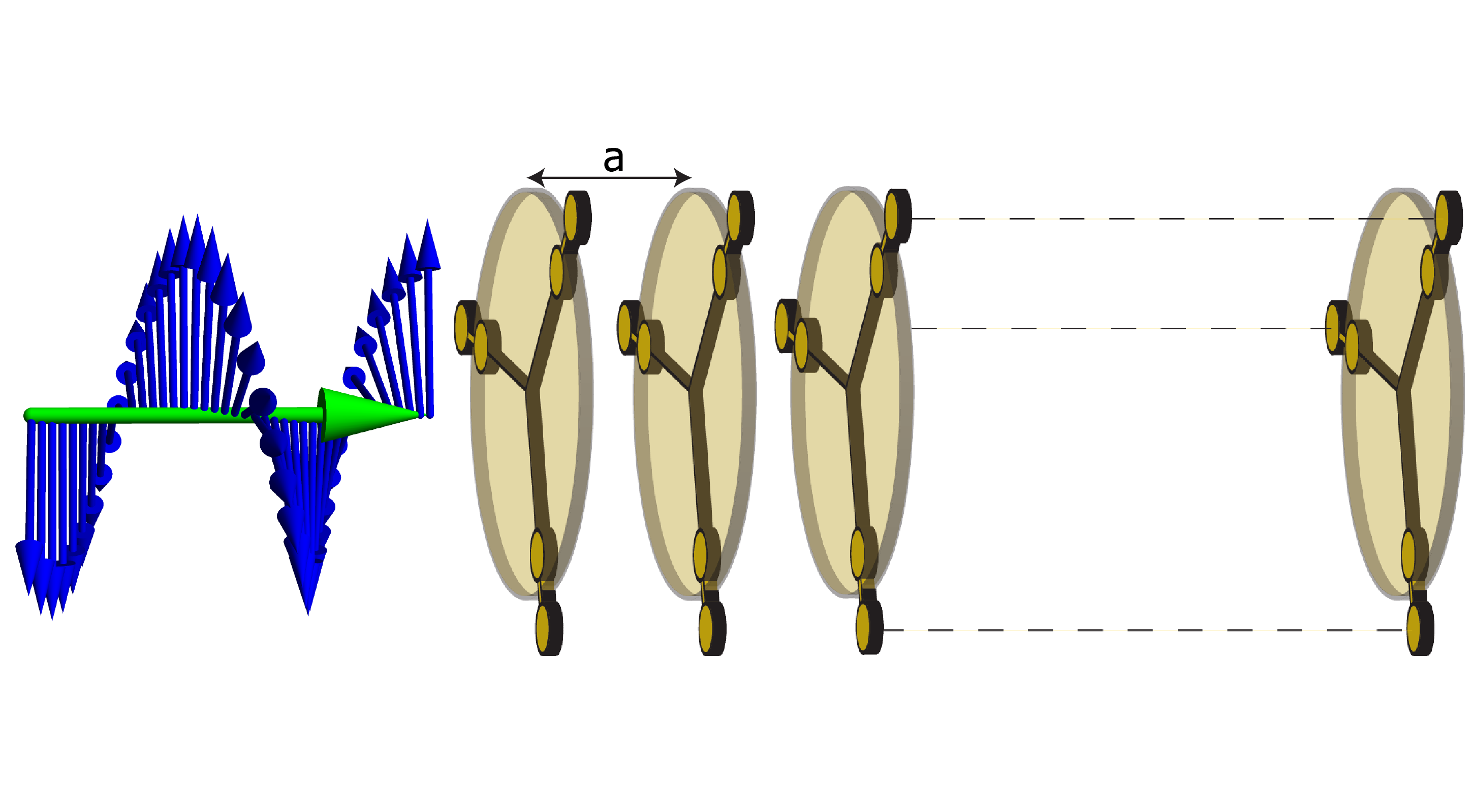}
\end{center}
\caption{\emph{20-site chain of three-arm $H_2$ sites.} A chain of 20 cofacial sites, separated by $a = 10.2$ Bohr and with arm radius $R = 3.78$ Bohr, is excited from the left by an optical vortex pulse. The resulting TE packet is shown in Fig. \ref{TXtrans}.}
\label{Chain_Triad}
\end{figure}

\begin{figure}[hptb]
\begin{center}
\includegraphics[width=0.45\textwidth]{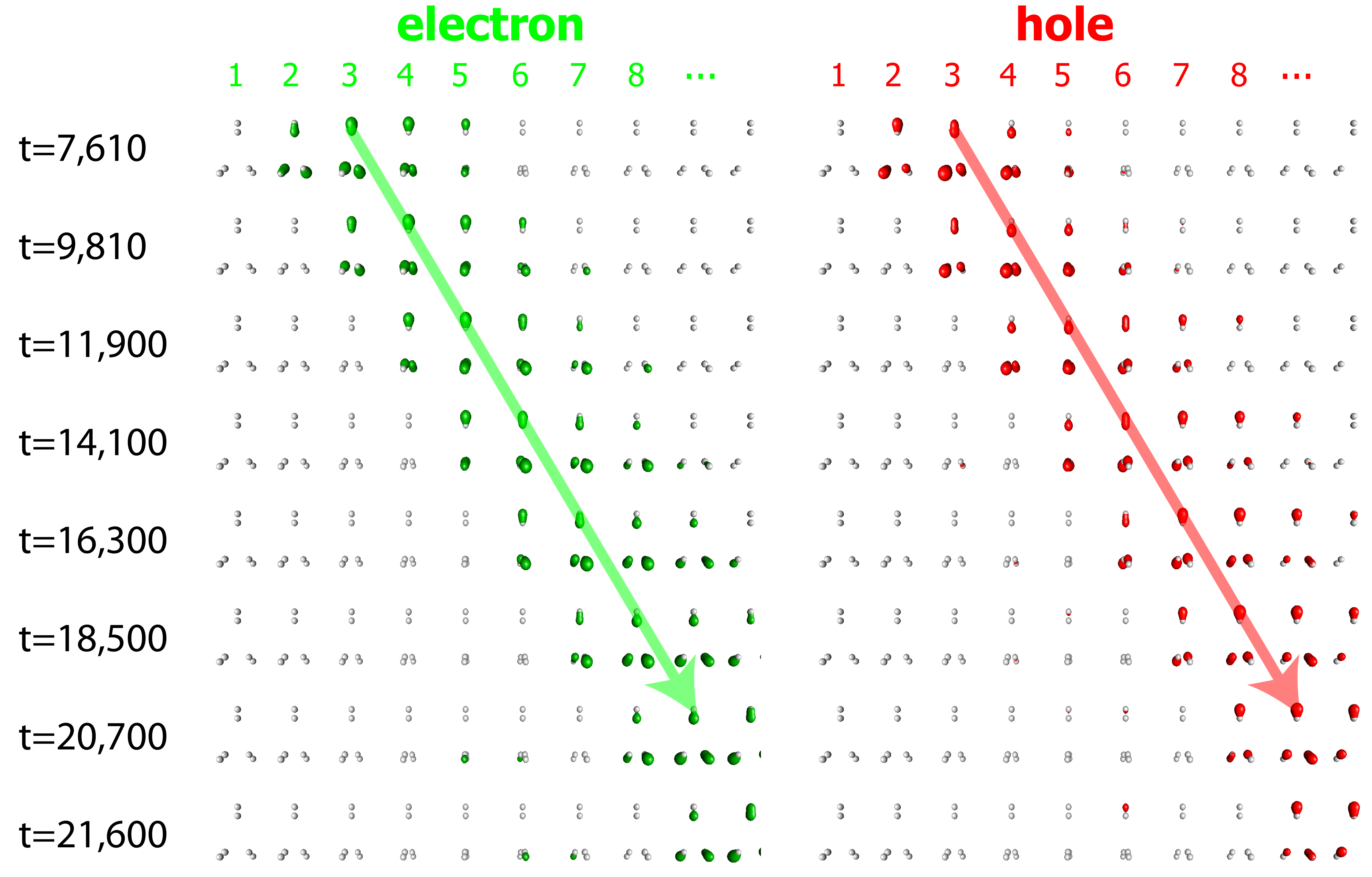}
\end{center}
\caption{\emph{TE transport in the 20-site, three-arm $H_2$ system.} Isosurface visualization of exciton transport with electron component (green) and hole component (red) densities of 0.005$/\mathrm{Bohr}^3$. Envelop parameters are $\{t_0 = 408, \tau=81.6, \omega=0.390, E_0 = 0.002\}$ in a.u.}
\label{TXtrans}
\end{figure}

With the site-to-site evolution of EAM now elucidated, attention can now be turned population dynamics on the 20-site, three-arm $H_2$ system depicted in Fig. \ref{Chain_Triad}. A laser with PAM = $1_p$ is applied to the left-most site of this chain, and the ensuing exciton dynamics is shown as a sequence of time slices in Fig. \ref{TXtrans}. Each time slice provides isosurfaces of electron (left, green) and hole (right, red) densities for the left-most ten sites on the chain. These isosurfaces are numerically constructed from attachment and detachment densities~\cite{AD, Zang_PRB_2017}. The speed of the energy transfer essentially constant, $5.48\, \mathrm{Bohr}/\mathrm{fs}$, demonstrating that even a simple Gaussian laser pulse can generate a TE packet with a well-defined speed. A tailored laser pulse, though, can be used to tune the packet shape and speed~\cite{Zang_PRB_2017}. Although the individual EAM components cannot be tracked on this larger chain, the results summarized in Figs \ref{DimerR2}(a) and \ref{DimerR1}(a) allow the conclusion that this packet is, indeed, a twisted exciton with EAM $= 1_e$.
%
%
%

%
%
%

\section{Conclusions}
Molecules with $C_N$ or $C_{Nh}$ symmetry support excitons with quantized states of excitonic angular momentum. Individual molecules can be viewed as mediators for changing the angular momentum of light, sidestepping the need to exploit nonlinear light-matter interactions based on higher-order susceptibilities~\cite{Zang_2017a}. Molecular chains composed of cofacial arrangements of these molecules support axial extended states that have both angular and linear momentum. This allows twisted exciton wave packets to be created, manipulated, and annihilated using laser pulses. Within the paraxial approximation, angular momentum is conserved in these light-matter interactions. Information encoded within the angular momentum of light can therefore be processed in excitonic form and subsequently re-emitted, forming the basis of opto-excitonic logic circuits. This adds an additional dimension of information content to generic Heisenberg spin chains~\cite{Bose_PRL_2003, Bose_ContempPhys_2007,Thompson_2016} that are often considered in association with quantum information processing~\cite{Wang_PRA_2012}. 

Both TB and RT-TD-DFT paradigms assumed that two-photon absorption can be used to excite isolated end sites without adversely affecting excitonic coupling to the rest of the system, as detailed in elsewhere~\cite{Zang_PRB_2017}. Exciton-phonon interactions were neglected but must be managed in order to maintain the requisite phase relationships in the coherence of state superpositions~\cite{Plenio_2008, Arago_AFM_2015}. Within this ansatz, an analytical relationship was derived that allows laser beams to be constructed so as to produce packets with a prescribed footprint, axial speed, and angular momentum. The methodology also shows how to use stimulated emission to transform the packet back into optical form. Such chain system have an excitonic band structure with individual bands associated with each value of EAM. The width of these bands can be controlled with the coupling between arms on adjacent sites, while the spacing between bands is dictated by the coupling between arms on a given site.

A TB setting was adopted to demonstrate that the laser pulses thus designed do generate wave packets with a well-defined angular momentum.  In particular, a computational methodology was developed to quantify the excitonic phase on each arm of every site, and this was applied to show that the angular momentum of light is transferred directly to excitonic form. It was also demonstrated that the angular momentum of these packets can be changed during transit using a second laser pulse.

More sophisticated RT-TD-DFT analyses, without many of the idealizations inherent in the simpler paradigm, were then used to elicit comparable results. TE packets were once again generated within a many-body setting with each arm capable of supporting a manifold of excited states. Because of the computationally intense nature of this approach, each arm was represented by a radially-oriented hydrogen dimer. An algorithm was developed that allows electron and hole populations to be estimated using attachment and detachment densities~\cite{AD}, allowing the exciton packet to be tracked as a function of time. A direct measure of the exciton phase was found to be computationally intractable for the 20-site chain, but RT-TD-DFT was used explicitly calculate phase evolution on the arms of a two-site triad of hydrogen molecules. This dimer analysis showed that excitonic angular momentum is transferred between sites. In both the TB and RT-TD-DFT settings, such resonant energy transfers between sites may be viewed as the result of the virtual emission and virtual absorption of a twisted radiation field.

A number of promising centrosymmetric molecules have been previously suggested in associated with single-molecule applications~\cite{Zang_2017a}, where a rudimentary analysis of triphenylphospine (TPP) suggested that it might be a good candidate for changing the angular momentum of light. Cofacial chains of such molecules should therefore be analyzed, along with other hydrogenated column five atoms. However, inert centrosymmetric scaffold structures are also attractive because they could be made with a large diameter and their periphery functionalized by chromophores sharing strong neighbor conjugation. This would move the active elements away from the dark axis of laser vortices. The larger systems, if conjugated with sufficient strength to support robust excitonic states, would allow higher angular momenta values to be transferred down molecular chains. Such scaffolds could also address the issue of how to fix the orientation and spacing of molecular constituents.

\section{Acknowledgements}

All calculations were carried out using the high performance computing resources provided by the Colorado School of Mines.


\end{document}